\documentclass[traditabstract]{aa}

\usepackage{graphicx}
\usepackage{txfonts}
\usepackage{natbib}
\usepackage{xspace}
\usepackage{xcolor}
\usepackage{hyperref}
\bibpunct{(}{)}{;}{a}{}{,}

\newcommand{\hc}{$H_0$}

\newcommand{\lcdm}{$\mathrm{\Lambda CDM}$\xspace}

\newcommand{\be}{\begin{equation}}
\newcommand{\ee}{\end{equation}}

\begin{document}

\title{Impact of the 3D source geometry on time-delay measurements of lensed type-Ia supernovae}

\author{
V.~Bonvin\inst{\ref{epfl}} \and
O.~Tihhonova\inst{\ref{epfl}} \and
M.~Millon\inst{\ref{epfl}} \and
J.~H-H.~Chan\inst{\ref{epfl}} \and
E.~Savary\inst{\ref{epfl}} \and
S.~Huber\inst{\ref{MPG}, \ref{TUM}} \and
F.~Courbin\inst{\ref{epfl}}
}

\institute{
Institute of Physics, Laboratory of Astrophysics, Ecole Polytechnique 
F\'ed\'erale de Lausanne (EPFL), Observatoire de Sauverny, 1290 Versoix, 
Switzerland \label{epfl}\goodbreak \and
Max Planck Institute for Astrophysics, Karl-Schwarzschild-Str.1, D-85740 Garching, Germany \label{MPG}\goodbreak \and
Physik-Department, Technische Universit\"at M\"unchen, 
James-Franck-Stra\ss{}e~1, 85748 Garching, Germany \label{TUM} \goodbreak
}

\date{\today}
\abstract{It has recently been proposed that gravitationally lensed type-Ia supernovae can provide microlensing-free time-delay measurements provided that the measurement is taken during the achromatic expansion phase of the explosion and that color light curves are used rather than single-band light curves. If verified, this would provide both precise and accurate time-delay measurements, making lensed type-Ia supernovae a new golden standard for time-delay cosmography. However, the 3D geometry of the expanding shell can introduce an additional bias that has not yet been fully explored. In this work, we present and discuss the impact of this effect on time-delay cosmography with lensed supernovae and find that on average it leads to a bias of a few tenths of a day for individual lensed systems. This is negligible in view of the cosmological time delays predicted for typical lensed type-Ia supernovae but not for the specific case of the recently discovered type-Ia supernova iPTF16geu, whose time delays are expected to be smaller than a day.}


\titlerunning{Microlensing time delay in SNeIa}
\maketitle

%======================
\section{Introduction}
%======================

Time-delay cosmography is a single-step method to measure the Hubble constant, \hc, independently of other techniques, such as cosmic microwave background observations \citep{Planck2016}, galaxy clustering and baryon acoustic oscillations \citep{DES2017}, or local distance ladder \citep{Freedman2012, Cao2017, Riess2018}. It can play an important role in assessing the validity of the standard cosmological model, that is, \lcdm \citep[e.g.,][]{Freedman2017}. The method consists in measuring the time delays between the multiple images of a source lensed by a foreground galaxy. If the source displays photometric variations, these will be seen at different epochs in each lensed image, allowing to measure a time delay between the different pairs of images. By reconstructing the mass profile of the lens galaxy one can turn the time delays into a so-called time-delay distance, which provides a direct measurement of \hc. The original idea of time-delay cosmography was proposed by \citet{Refsdal1964}, who suggested to use lensed supernovae as variable sources. For a long time, however, lensed quasars were used instead as the probability of observing lensed supernovae was very low. Time-delay measurement in the lensed quasar, which is the core of the COSMOGRAIL program \citep[e.g.,][]{Courbin2005,Tewes2013b,Eulaers2013,Courbin2017, Rathna2013}, must be both precise and accurate, as any error on the time delays propagates linearly to \hc. COSMOGRAIL time delays were used to estimate \hc\ in the context of the H0LiCOW program, leading so far to a precision of 3.8\% on \hc, including systematics \citep{Bonvin2017, Sluse2017, Suyu2017, Rusu2017, Tihhonova2017, Wong2017}.

While the prospects to further improve the precision on \hc\ are excellent \citep[e.g.,][]{Treu2016, Suyu2017, Suyu2018, deGrijs2017}, there still exists a number of difficulties to overcome. Among them is the presence of compact objects in the lens galaxy, which act as secondary lenses (or microlenses), producing photometric variations in the light curves of lensed images \citep[e.g.,][]{Tewes2013a}. In addition, microlensing can introduce an extra time delay that is unrelated to cosmology and that may therefore bias it. For quasars, according to the ``lamp-post" model \citep[][TK18]{Tie2017}, this effect is related to the way luminosity variations propagate across the accretion disk, which has a finite size.

The two discoveries of lensed supernovae \citep{Kelly2015, Goobar2017, Grillo2018} and the prospects for discovering many more with future large-sky surveys \citep{Oguri2010} has caused a renewed interest for these objects. Lensed type-Ia supernova (SNeIa) have in principle multiple advantages over lensed quasars: their standard candle nature can help to break the degeneracies in lens models such as the mass sheet transformation and the source position transformation \citep[e.g.,][]{Schneider2013, Schneider2014}, and time-delay measurements can benefit from the knowledge of a known template or family of templates for the light curves. In addition, \citet[][hereafter G18]{Goldstein2018} studied the impact of microlensing on lensed SNeIa and found that during the rest-frame weeks after the explosion, in the so-called achromatic growing phase, color curves are free of microlensing. This makes the time-delay measurements in SNeIa all the more accurate than in quasars. Based on this, \citet[][hereafter FM18]{FoxleyMarrable2018} forecast a $0.5\%$ precision measurement for \hc\ by using a specific subset of lensed SNeIa from the Large Synoptic Survey Telescope (LSST), following predictions by \citet{Goldstein2017}. However, reaching such a precision assumes that no systematic biases of the same order are affecting the individual measurements. 

In this work, we report such a systematic effect arising due to the 3D geometry of the expanding shell of SNeIa. In projection on the plane of the sky, photons emitted at the center of the shell reach the observer earlier than photons emitted at the edges. This delay, caused purely by the geometry of SNeIa, skews the observed light curves in all bands, and introduces a bias both in time and in magnitude with respect to a point source. The delay scales with the angular size of the SNeIa, and in the presence of microlensing differs between the lensed images, as microlensing magnifies the surface brightness profile of each lensed image in a different way.

In the following, we introduce terminology in order to avoid potential confusion, and describe a simple formalism to compute microlensing time delay as a function of SNeIa and lens galaxy parameters. Our results are illustrated by estimating the bias on time delays in mock SNeIa light curves. 

%====================
\section{Terminology}
%====================

The present work can be seen as an extension of the TK18 principles to the case of lensed SNeIa. There is, however, some confusion in the literature as to the precise nature of microlensing time delay. In order to remove any ambiguity we propose to adopt the following terminology.

\begin{itemize}
\item {\bf Cosmological time delay:} the delay produced by the smooth mass profile of the lensing galaxy. This ranges from days to months and depends only on cosmology and on the mass distribution (including sub-haloes of dark matter) in the lens.

\item {\bf Observed time delay:} often simply referred to as the \emph{time delay}, that is, the delay directly measured from light curves and usually quoted in publications, whatever be the additional factors affecting it independently of cosmology. 

\item {\bf Geometrical time delay:} the delay that corresponds to the distortion of the light curves due to the two-dimensional (2D) or 3D geometry of the source as photons emitted from different regions reach the observer at different times \citep[see e.g.,][]{Lucy2005}. We note that this effect occurs whether or not the source is lensed.

\item {\bf Microlensing magnification:} a shift in magnitude potentially variable in time induced by microlensing on the intrinsic light curve of the source. This can lead to distortions of the light curves in the lensed images and impact the way time delays are measured. It can also be chromatic, depending on the spatial energy profile of the source, but is not to be mistaken for the microlensing time delay.

\item {\bf Microlensing time delay:} the additional effect induced by the reweighting of the geometrical time delay by the microlensing pattern affecting the extended source. We note that this definition slightly differs from TK18, where their microlensing time delay was the combination of our geometrical and microlensing time delays.

\item {\bf Excess of time delay:} any time delay applied to a single lensed image. Excess time delays cannot be measured directly, whereas the difference of excess delay, usually simply referred to as time delay, can be measured between a pair of lensed images.

\end{itemize}

In the absence of microlensing, if the source is a perfect point source, the observed delay corresponds to the cosmological delay. If the source is extended, an excess of geometrical time delay can take place. As this geometrical time delay is the same in each lensed image, it cancels out between pairs of lensed images. In the presence of microlensing, however, this excess of geometrical delay is weighted differently in each lensed image as they are affected by different networks of micro-caustics. This creates a differential excess of microlensing time delay between the lensed images, which biases the observed time delay as compared to the delay of interest, i.e., the cosmological time delay.

%=================================================
\section{Geometrical and microlensing time delays}
%=================================================

The derivations presented in this section are in essence analogous to the lamp-post model for quasars used by TK18, although there are differences with TK18 leading us to use a slightly different  formalism.

\begin{figure}[t!]
    \centering
    \includegraphics[scale=0.6]{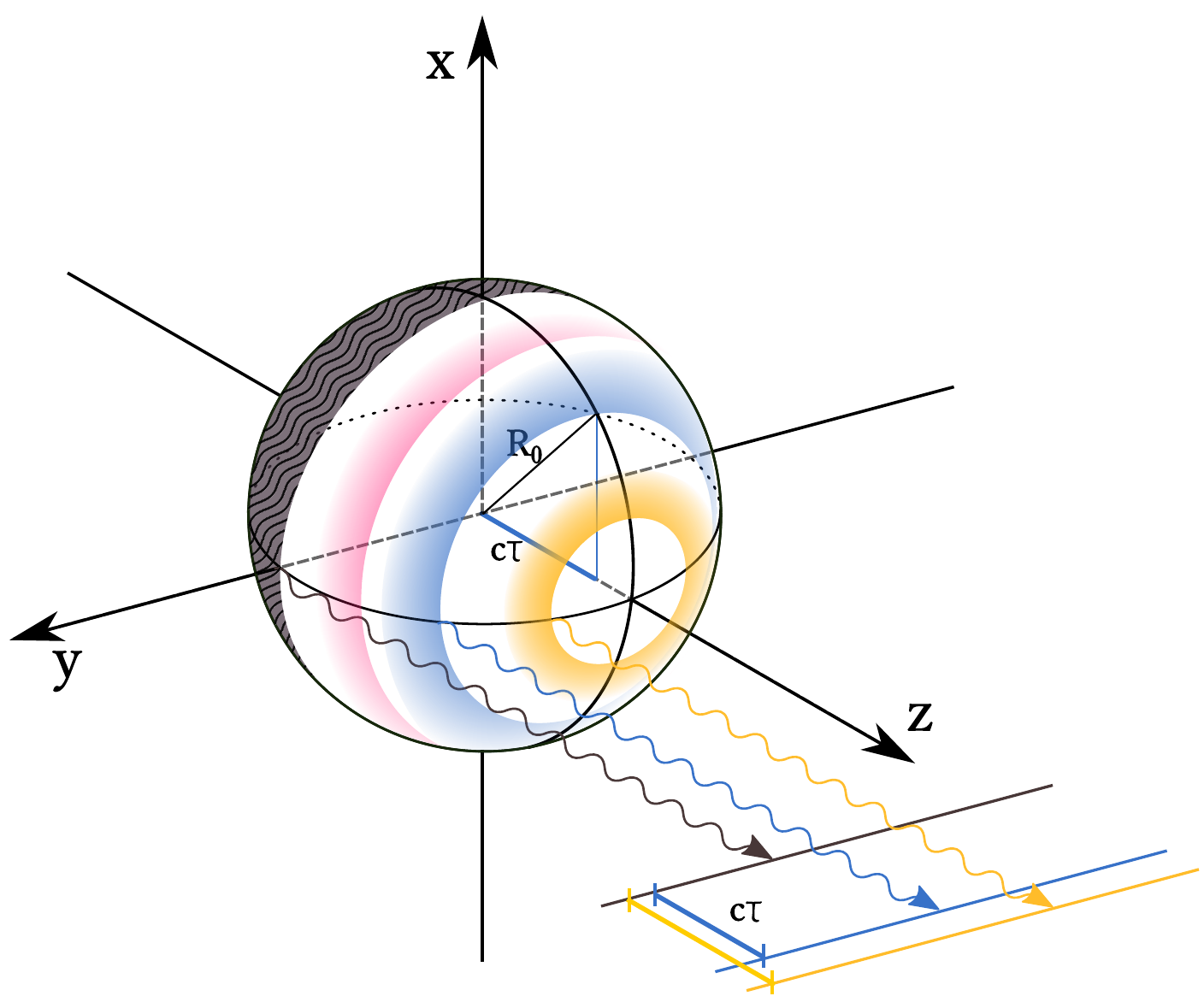}
        \caption{Illustration of the geometry of the problem. At a given time $t$, a shell of radius $R_0 (t)$ emits photons towards the observer (along the z-axis). Different regions of different z coordinates (illustrated here with three colored circles) are located at different distances from the observer. Thus, for all the photons that reach the observer at a given time, there is a corresponding geometrical delay $\tau$ of their emission time.}
    \label{fig:sheme}
\end{figure}

Our model assumes optical thickness during the first weeks after explosion and spatially constant surface brightness $S$ (i.e., no limb darkening). The emission profile is thus modeled as a spherically symmetric expanding shell, as illustrated in Fig.~\ref{fig:sheme}, which we project onto a disk. Considering the disk projection, the photons that reach the observer simultaneously are emitted with a delay depending on the distance to the disk center. From the point of view of the observer, the surface brightness of the disk can be written as $S=S(x,y,t+\tau)$. The emission delay $\tau(x,y,t)$ of the light emitted from any point of the disk is: 

 \be \label{eq:geometrical_delay}
 \tau (x,y,t) = \frac{\sqrt{R_0(t)^2-x^2-y^2}}{c}, 
 \ee
where $R_0(t)$ is the radius of the supernova at observer time $t$ after the explosion. The delays are relative to the plane perpendicular to the line of sight and passing through the center of the supernova, that is, $z=0$. Following this definition, $\tau = 0$ corresponds to the emission from the edges of the disk whereas $\tau = R_0(t)/c$ corresponds to the emission from the center. The value of $\tau$ is always positive, and corresponds to the delay of the time of emission of the wavefronts that reach the observer simultaneously. Before going further, there are two things to be noted. First, Eq.~\ref{eq:geometrical_delay} assumes a nonrelativistic expansion of the shell; in the relativistic scenario $R_0$ has to be evaluated not at observer time $t$ but at the time of emission which varies with $(x, y)$\footnote{We performed our analysis including the relativistic corrections of the first order and found no significant deviation from the nonrelativistic case.}. Second, the spatially constant surface brightness assumption means that for any two sets of time-coordinates $(x_1, y_1, t_1)$ and $(x_2, y_2, t_2)$ respecting $t_1+\tau(x_1, y_1, t_1) = t_2+\tau(x_2, y_2, t_2)$, the surface brightness is the same, that is, $S(x_1, y_1, t_1+\tau(x_1, y_1, t_1)) = S(x_2, y_2, t_2+\tau(x_2, y_2, t_2))$. The surface-brightness profile can therefore be fully characterized by the time evolution on a single set of coordinates, for example $S(0, 0, t+\tau(0, 0, t))$. We nevertheless keep the spatial coordinates in the surface brightness expression to be explicit.

The lensing galaxy located in front of the source magnifies the source intrinsic luminosity. The amplitude of the magnification depends on the alignment between the source, the lens, and the observer. Both the smooth mass profile of the galaxy (the macrolens) and individual stars in it (the microlenses) contribute to the total magnification. To a given lens system composed of macro- and microlenses corresponds a magnification pattern in the source plane $M(x,y)$ \citep{Wambsganss1992} that can be separated into macro and micro magnification patterns, such that $M(x,y)=M_{macro}\cdot M_{micro}(x,y)$. The expression of the total observed luminosity $L(t)$ of a lensed SNeIa image can thus be computed by integrating the surface brightness of the disk at the time of the emission $S(x,y,t+\tau)$ multiplied by the magnification pattern\footnote{We express the surface brightness $S$ as function of the spatial coordinates $(x,y)$ to emphasize that it is differentially weighted by $M(x,y)$}:
  \be\label{eq:flux}
  L(t) = \int_{-x_{\mathrm{lim}}}^{x_{\mathrm{lim}}} \int_{-y_{\mathrm{lim}}}^{y_{\mathrm{lim}}} S(x,y,t+\tau)\, M(x,y)\, dx\, dy, 
  \ee
with $x_{\mathrm{lim}}=R_0(t)$ and $y_{\mathrm{lim}}=\sqrt{R_0(t)^2 - x^2}$.
Even in the absence of microlensing ($M(x,y)=M_{macro}$, i.e., magnification comes from the macro model only), the observed light curves are distorted if we consider the supernova as a 3D source, because we account for the differential travel time of the photons. This distortion is what we call the excess of geometrical time delay. In the presence of microlensing, one immediately notices that different regions of the SNeIa surface are differentially magnified by the microlensing pattern. This re-weighting leads to an additional contribution with respect to the nonmicrolensed case, the so-called excess of microlensing time delay. 

The combined mean excess of geometrical and microlensing time delay $\langle\tau^{\mathrm{gm}}\rangle(t)$ can be computed simply by weighting the emission delay $\tau$ in Eq.~\ref{eq:geometrical_delay} with the observed supernova luminosity in Eq.~\ref{eq:flux}:
 \be\label{eq:mldelay}
 \langle \tau^{\mathrm{gm}} \rangle (t) =  \frac{1}{L(t)} \int_{-x_{\mathrm{lim}}}^{x_{\mathrm{lim}}} \int_{-y_{\mathrm{lim}}}^{y_{\mathrm{lim}}} S(x,y,t+\tau)\, M(x,y) \, \tau \, dx \,dy.  
 \ee
 
The mean excess of geometrical time delay alone $\langle \tau^{\mathrm{g}} \rangle$ can be computed by setting $M(x,y)=M_{macro}$. As $M_{macro}$ is a constant, it can be extracted from the integral in Eqs.~\ref{eq:mldelay} and~\ref{eq:flux}, and cancels out in Eq.~\ref{eq:mldelay}. This shows that the magnification from the macromodel has no impact on the excess of geometrical and microlensing time delays. The excess of microlensing time delay alone $\tau^{\mathrm{m}}$ can thus be obtained by simply subtracting the excess of geometrical time delay from Eq.~\ref{eq:mldelay}. When working with a source located at redshift $z_s$, the light curves as seen by the observer are stretched by a factor of $(1+z_s)$ with respect to the light curves in the source referential, and all the delays in the observer referential correspond to the delays in the source referential multiplied by $(1+z_s)$. In this work, we compute all the quantities in the source referential and rescale (multiply) the light curves (time delays) by $1+z_s$ when illustrating our results for specific cases.

%=====================
\section{Results and discussion}
%=====================

\begin{figure*}[h]
    \centering
    \begin{minipage}[l]{0.41\textwidth}
    \includegraphics[width=0.99\linewidth]{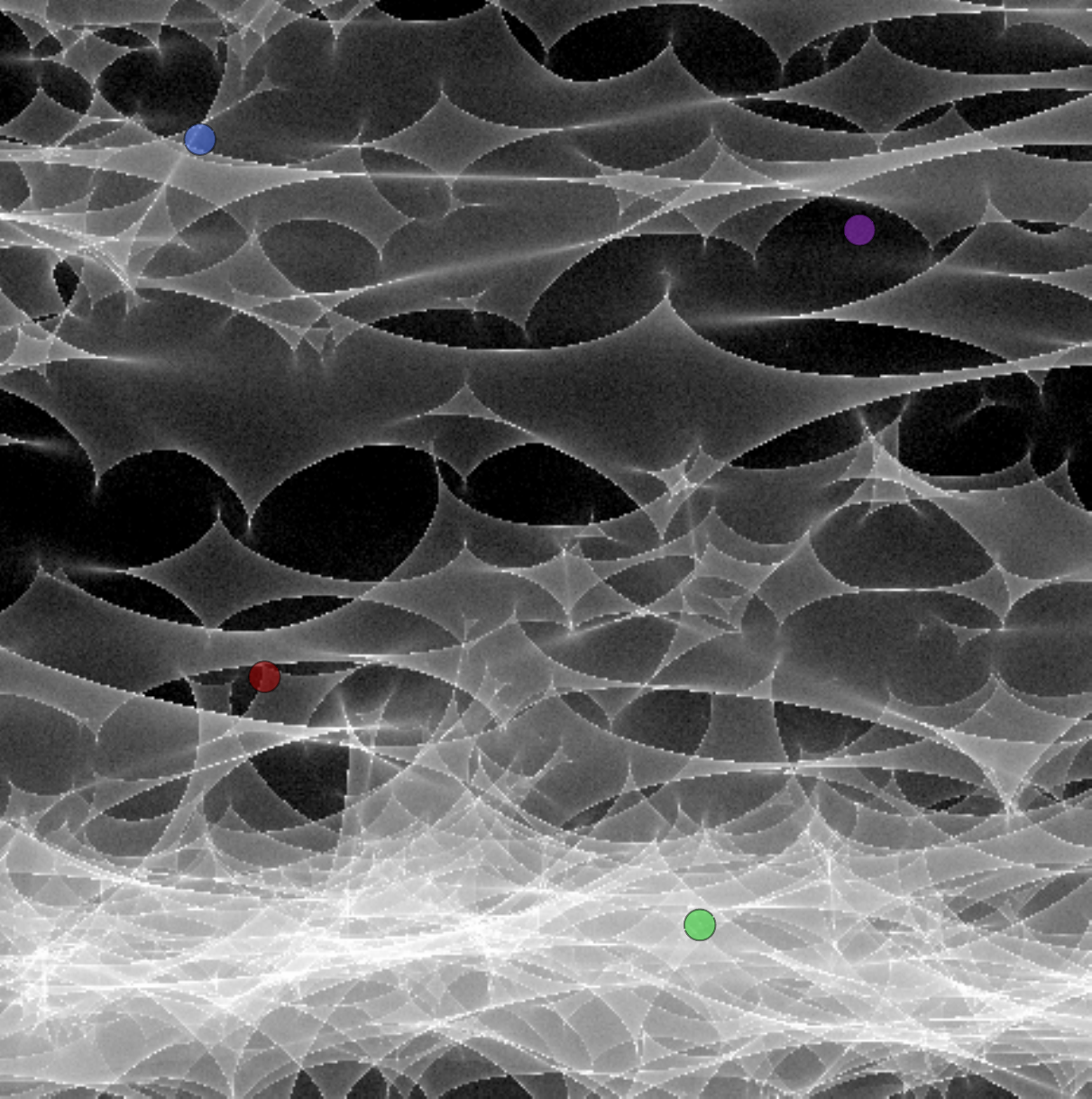}
    \end{minipage}
    \begin{minipage}[r]{0.58\textwidth}
    \includegraphics[width=0.99\linewidth]{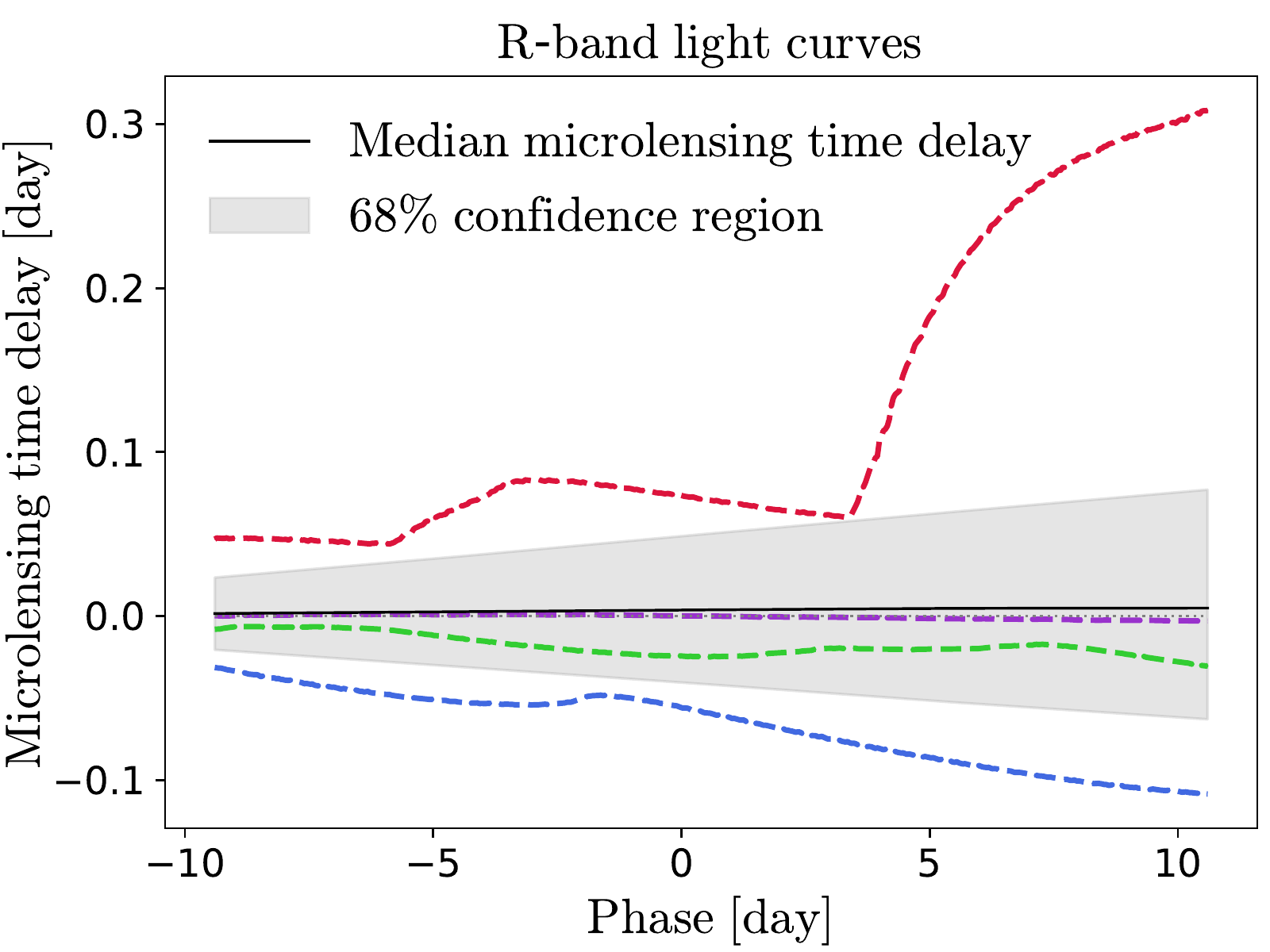}
    \end{minipage}
    \caption{\textit{Left:} Zoom-in on the microlensing magnification map for image D of iPTF16geu, with four different source positions indicated with colored disks. The size of the disks correspond to the projected physical size of the supernova 10 days after peak luminosity. White regions in the map correspond to higher magnification. \textit{Right:} Evolution of the microlensing time delay over time for the D image of iPTF16geu in our toy model. Phase 0 corresponds to the peak luminosity in R-band of the unlensed point-source template. The solid black line indicates the median delay and the shaded gray envelope covers the 68$\%$ confidence region resulting from centering the source on all the possible positions in the magnification map. The colored dashed curves correspond to sources at the location of the colored circles in the left panel. On average, the amplitude of the microlensing time delay is not larger than $0.1$ days, but can become much higher in extreme cases.}
    \label{fig:mltd_evol}
\end{figure*}

We illustrate the excess of geometrical and microlensing time delays using a toy model for SNeIa. We choose as a source of surface-brightness time evolution $S(0, 0, t)$ the nearby unlensed supernova SN2011fe \citep{Nugent2011} in order to avoid additional lensing effects in the light curves. We use the expansion velocity of the photosphere given by \citet{Goobar2017}, and the flux modeled using the SNcosmo \citep{SNcosmo} implementation of SALT2 \citep{Guy2007} based on the observations of \citet{Pereira2013}. We then rescale the flux to the redshift of the supernova, $z_{SN}=0.409$, to match a source-lens configuration of the lensed supernovae iPTF16geu \citep{Goobar2017} - the only resolved lensed SNeIa discovered to date. Following G18, we assume the same surface-brightness profile $S(x,y,t)/S(0,0,t)$ in all bands, leading to the so-called achromatic expansion phase. To keep this assumption realistic, we model the SNeIa only up to $\sim$10 days after the luminosity peak in R-band. We generate the magnification maps based on the iPTF16geu lens at the positions of the multiple images using the parameters from Table~1 by FM18, that partly relies on the {\tt GLAFIC SIE} model of \citet{More2017}. We use the inverse ray-shooting technique \citep{Wambsganss1992} implemented on graphics processing units \citep[{\tt GPU-D}][]{Vernardos2014}, following the same formalism as in \citet{Bonvin2018}. The size of the magnification maps is $20\times20$ Einstein radii, with a mean stellar mass of $M_{\star}=0.3M_{\odot}$. The microlensing magnification maps are obtained by dividing the total magnification maps by the magnification obtained in the absence of microlensing. The left panel of Fig.~\ref{fig:mltd_evol} shows a zoom-in (approx. one quarter of the total map) into our realization of the microlensing map $M_{micro}(x,y)$ for image D of iPTF16geu. In what follows, unless otherwise specified, our results always refer to image D. This choice is purely aesthetic, but we assessed that our conclusions remain valid for the other images.

\subsection{Impact of microlensing time delay on the measured delay}

The first practical application of our toy model is to compute the average amplitude of the microlensing time delay. To do so, we first select the filter we are interested in that will determine the choice of source brightness time evolution $S(0, 0, t)$. From Eq.~\ref{eq:mldelay} we can compute the evolution of the mean excess of geometrical and microlensing time delay over time, with the source centered at a given position on the microlensing map and growing in size over time. Centering the source on all the possible positions in the microlensing map, one obtains the probability distribution of the mean excess of geometrical and microlensing time delay at a given time $\langle \tau^{\mathrm{gm}} \rangle(t)$, for a chosen microlensing pattern $M_{micro}(x,y)$ statistically representative of the microlensing at the lensed images position. One can also compute the mean excess of geometrical time delay alone $\langle \tau^{\mathrm{g}} \rangle(t)$ by assuming no microlensing, that is, $M(x,y)=M_{macro}$. Subtracting the geometrical contribution to $\langle \tau^{\mathrm{gm}} \rangle(t)$ gives the probability density of the mean excess of microlensing time delay $\langle \tau^{\mathrm{m}} \rangle(t)$. The right panel of Fig.~\ref{fig:mltd_evol} presents the time evolution of the median excess of microlensing time delay and the 68$\%$ confidence region envelope for image D of iPTF16geu using the R-band source surface brightness time evolution. The envelope size increases over time due to the expansion of the supernova thus enhancing both the geometrical and microlensing time delays.

%The colored dashed curves represent four specific configurations, corresponding to different positions in the magnification map on the left panel.

%We also observe that the median microlensing time delay is positive and increasing with time. This is in agreement with TK18, which also noticed a positive bias. We interpret this result as the probability of magnifying the outer part of the supernovae, which have a positive excess of microlensing time delay being higher as magnifying the central part. This is a purely geometrical effect, as the positive excess regions cover a bigger area on the sky and have therefore a higher probability to be magnified. 

The distribution of excess of microlensing time delay at a given time for our toy model corresponds to a vertical slice in the microlensing time-delay evolution presented in the right panel of Fig.~\ref{fig:mltd_evol}. It can be noted from Eq.~\ref{eq:mldelay} that the choice of the filter will impact the microlensing time delay, depending on how the surface-brightness profile $S(x, y, t+\tau)$ varies over time. In practice, the time gradient of the surface brightness is small compared to $\tau$, that is, $S(x, y, t+\tau)$ is nearly constant at a given time $t$ but variable $(x, y)$, and therefore the choice of surface-brightness time evolution $S(0, 0, t)$ plays a negligible role in the integral. This translates into a very weak dependency on the chosen filter (a maximum difference of the order of $10^{-6}$ days whether we chose a U, B, V, R or I filter). We report in the first row of Table~\ref{tab:tab} the 16th, 50th and 84th percentiles of distributions for individual lensed images (the excess) as well as for image pairs (difference of excesses) computed 10 days after the peak luminosity of the R-band point-source template, and emphasize that these results, at the precision quoted, are \textbf{independent of the choice of filter}. We also report the microlensing time-delay contribution to the observed time delays for different pairs of images, obtained by cross-correlating the individual distributions of excess of microlensing time delay.

For the majority of configurations explored here, the bias from microlensing time delay is below $0.1$ days, as highlighted in the first row of Table~\ref{tab:tab}. We also find that microlensing time delays become more important in specific cases. Looking at the probability distribution of microlensing time delay at $t=10$ days, we find that there is a probability of $\sim1/10$ that the microlensing time delay for the AB image pair exceeds 0.12 days, $\sim1/100$ that it exceeds 0.3 days, and $\sim1/1000$ that it exceeds 0.45 days. Consequently, as long as one does not aim for a precision on time-delay measurements smaller than a day, the contribution of microlensing time delay to the total error budget remains small ($\leq10\%$). Using a large enough sample of lensed SNeIa, the microlensing time delays affecting the measured time delay of the individual lensed systems average out to the median values presented in Table~\ref{tab:tab}, which, although nonzero, are one order of magnitude smaller than the error on the individual systems.

\begin{table*}

\caption{Predicted microlensing time delay distributions for all individual images and pairs of images of our toy model.}

\centering
\begin{tabular}{l c c c c c c c c c c}
\hline\hline

 & A [day] & B & C & D & AB & AC & AD & BC & BD & CD \\ \hline

\raisebox{-3.5pt}{full}
& \raisebox{-3.5pt}{$.005_{-.088}^{+.092}$}
& \raisebox{-3.5pt}{$.005_{-.080}^{+.080}$}
& \raisebox{-3.5pt}{$.005_{-.076}^{+.080}$}
& \raisebox{-3.5pt}{$.001_{-.068}^{+.072}$}
& \raisebox{-3.5pt}{$-.003_{-.128}^{+.124}$}
& \raisebox{-3.5pt}{$-.003_{-.128}^{+.124}$}
& \raisebox{-3.5pt}{$-.007_{-.120}^{+.116}$}
& \raisebox{-3.5pt}{$.001_{-.120}^{+.120}$}
& \raisebox{-3.5pt}{$-.007_{-.108}^{+.112}$}
& \raisebox{-3.5pt}{$-.007_{-.112}^{+.108}$} \\

\raisebox{-3.5pt}{excess}
& \raisebox{-3.5pt}{$-.007_{-.076}^{+.076}$}
& \raisebox{-3.5pt}{$-.019_{-.072}^{+.092}$}
& \raisebox{-3.5pt}{$-.007_{-.088}^{+.104}$}
& \raisebox{-3.5pt}{$-.007_{-.072}^{+.072}$}
& \raisebox{-3.5pt}{$-.007_{-.116}^{+.112}$}
& \raisebox{-3.5pt}{$.001_{-.124}^{+.124}$}
& \raisebox{-3.5pt}{$-.003_{-.104}^{+.108}$}
& \raisebox{-3.5pt}{$.005_{-.128}^{+.132}$}
& \raisebox{-3.5pt}{$.005_{-.116}^{+.108}$}
& \raisebox{-3.5pt}{$-.007_{-.124}^{+.124}$} \\

\vspace*{-0.25cm}
\\
\hline
\end{tabular}
\tablefoot{The quoted values are the 50th, 16th and 84th percentiles of the excess and difference for the excess of microlensing time delay distributions. The distributions are computed 10 days after the peak luminosity. The first row shows the values obtained when centering the source on all possible positions in the respective microlensing map, or cross-correlating the individual images distributions in the case of image pairs. The second row shows the values obtained when reweighing the maps by the observed flux excesses in iPTF16geu (see text). Note that we used the unlensed R-band point source template to model the surface brightness time evolution at the center of the disk (i.e. $S(0, 0, t)$), but choosing another band or even a color curve does not affect the results.}
\label{tab:tab}
\end{table*}

\subsection{The specific case of iPTF16geu}

\citet{More2017} inferred by studying the flux anomaly of iPTF16geu that this SNeIa undergoes significant microlensing magnification. However, both \citet{Yahalomi2017} and FM18 found that microlensing alone is very unlikely to be the only source of the extreme magnification of image A. In this section, we first test whether our toy model, which takes into account the geometry of SNeIa, confirms this statement. Then, we explore if the observed magnification can help to constrain the microlensing time delay. We conclude by commenting on how microlensing time delay affects the precision of the measured time delays in iPTF16geu.

Assuming given macro and micro lens models for iPTF16geu (in the present case, the {\tt GLAFIC SIE} model of \citet{More2017} with stellar mass fraction of FM18) and taking advantage of the standard candle nature of the source, we compute the predicted macro and micro magnifications for lensed images at a given time and compare them to the observed magnification presented in \citet{More2017}. Removing the magnification predicted for the macro model, we are left with the pure microlensing contribution and can assess how likely it is that the observed magnification can be reproduced by microlensing only. Our results differ from the works cited above as we use a source intensity profile accounting for its 3D nature, whereas FM18 use a uniform disk\footnote{We are able to reproduce very closely Fig.~8 of FM18 by using similar source parameters, i.e., spatially constant profile brightness and a shell expansion velocity of $10'000$ km/s.} and \citet{Yahalomi2017} use a point source. 

\begin{figure}[h]
    \centering
    \includegraphics[width=0.49\textwidth]{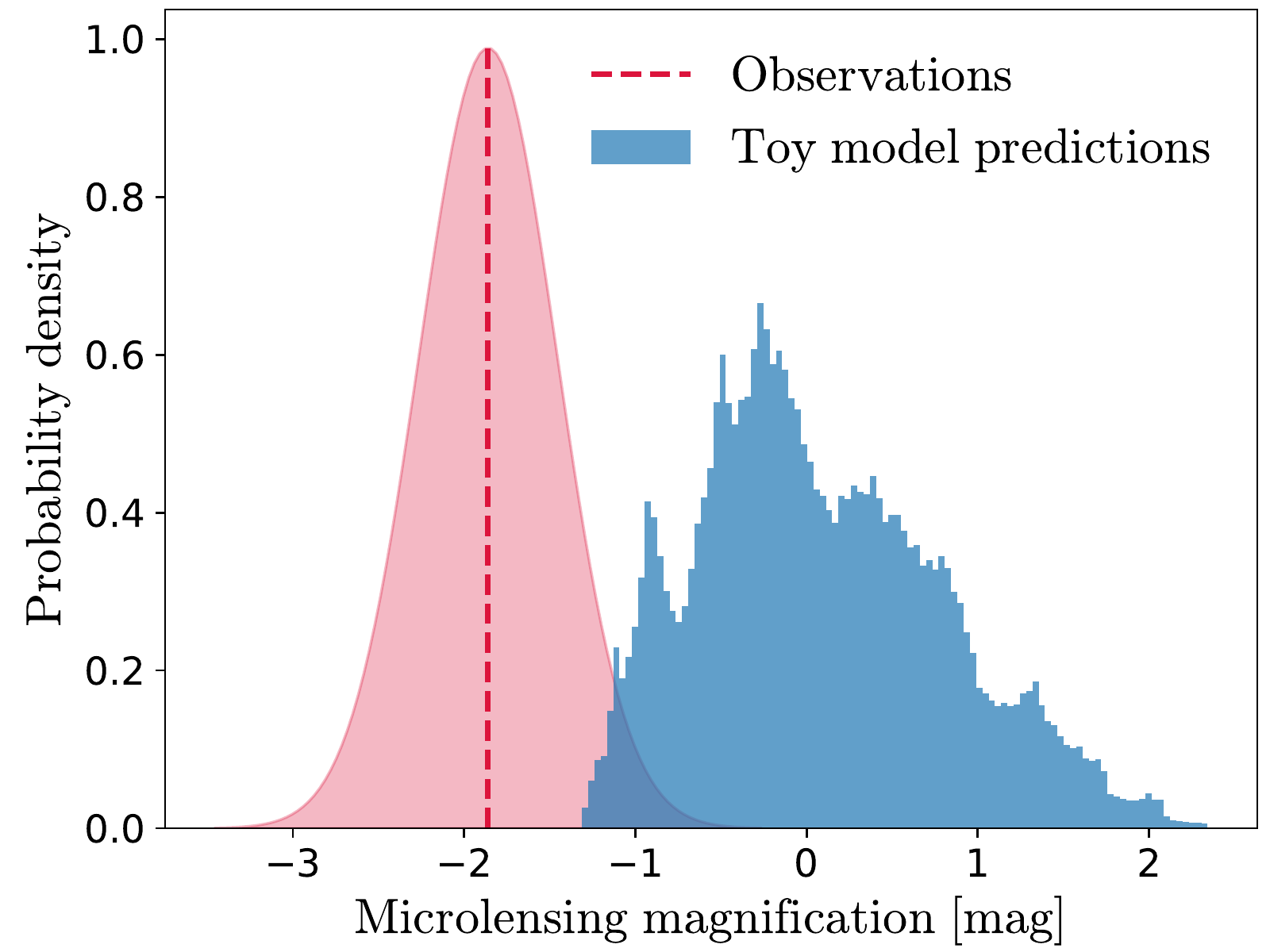}
    \caption{In blue, distribution of the microlensing magnification for image A of our toy model 25 days after peak luminosity. In red, Gaussian distribution centered on the observed excess of magnification minus the predicted macro model magnification of \citet{More2017}. The Gaussian distribution can be used as a prior to constrain the regions in the magnification maps where the source is more likely to be located.}
    \label{fig:mlmag_A}
\end{figure}

We present in Fig.~\ref{fig:mlmag_A} the probability density function of the magnification due to microlensing only, $\sim35$ days after the luminosity peak in the observer frame ($\sim25$ days in the iPTF16geu rest frame\footnote{At this point, SNeIa are not optically thick anymore, yet this should not significantly affect the predicted excess of magnification as long as the source size is correctly represented.}) for image A. We note that our toy model has a faster shell expansion than in FM18. We also plot a Gaussian distribution centered on the observed magnitude from \citet{Goobar2017} from which we subtract the {\tt GLAFIC SIE} predicted macro-model magnification from \citet{More2017}. The 1$\sigma$ uncertainties of the Gaussian distribution are $\Delta m=0.45$, following the error of the macro model magnification predictions. We see that accounting for the modeling error broadens the probability density function of the observations sufficiently to overlap with the predicted micro magnification.

We use this overlap to further constrain the possible values of the excess of microlensing time delay, using the observations as a prior for the predicted magnification map. We associate a weight $w(x,y)$ to each pixel of the map, according to how the micro magnification predicted at phase $t=25$ days for a source centered on this pixel agrees with the prior. We then recompute the microlensing time-delay distributions at any time by changing $M(x,y) \rightarrow w(x,y)\times M(x,y)$  in Eq.~\ref{eq:mldelay}. %Weights $w(x,y)$ are defined as the values of the Gaussian distribution at the corresponding predicted magnification at position $(x,y)$. 

The value of the weighted microlensing time delay distributions $10$ days after peak luminosity are presented in the second row of Table~\ref{tab:tab}. It is interesting to note that accounting for the observations does not significantly affect the predicted microlensing time delays. This illustrates well the fact that it is not the absolute amount of magnification that drives the amplitude of the microlensing time delay, but rather the spatial variations of the magnification across the source. Indeed, a constant microlensing magnification in Eq.~\ref{eq:mldelay} is equivalent to no microlensing in terms of microlensing time delay. This is also illustrated by the microlensing time-delay curves in Fig.~\ref{fig:mltd_evol}: the red and blue curves show higher values for the  microlensing time delay than the green curve, even though they correspond to regions in the magnification map where the total magnification is smaller.

We conclude this section by commenting on the impact of microlensing time delay on the use of iPTF16geu for time-delay cosmography. The model predictions of the cosmological time delays explored in \citet{More2017} are  $\sim0.5$ days. In this case, microlensing time delay introduces a significant discrepancy between the observed and cosmological time delays (a difference $>20\%$) that translates directly into $H_0$, potentially limiting the use of iPTF16geu for time-delay cosmography.

\begin{figure*}[h]
    \centering
    \begin{minipage}[l]{0.49\textwidth}
    \includegraphics[scale=0.49]{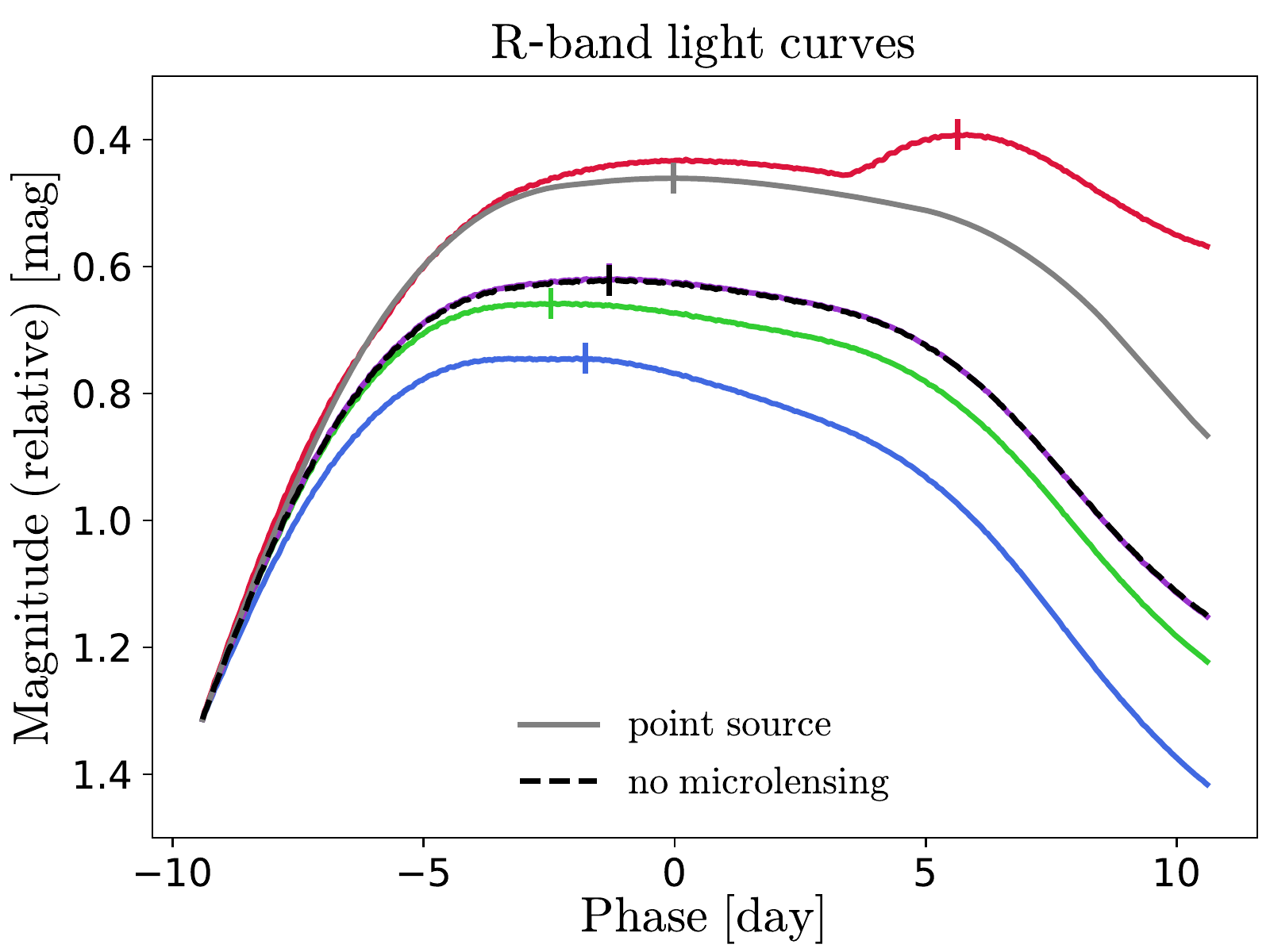}
    \end{minipage}
    \begin{minipage}[r]{0.49\textwidth}
    \includegraphics[scale=0.49]{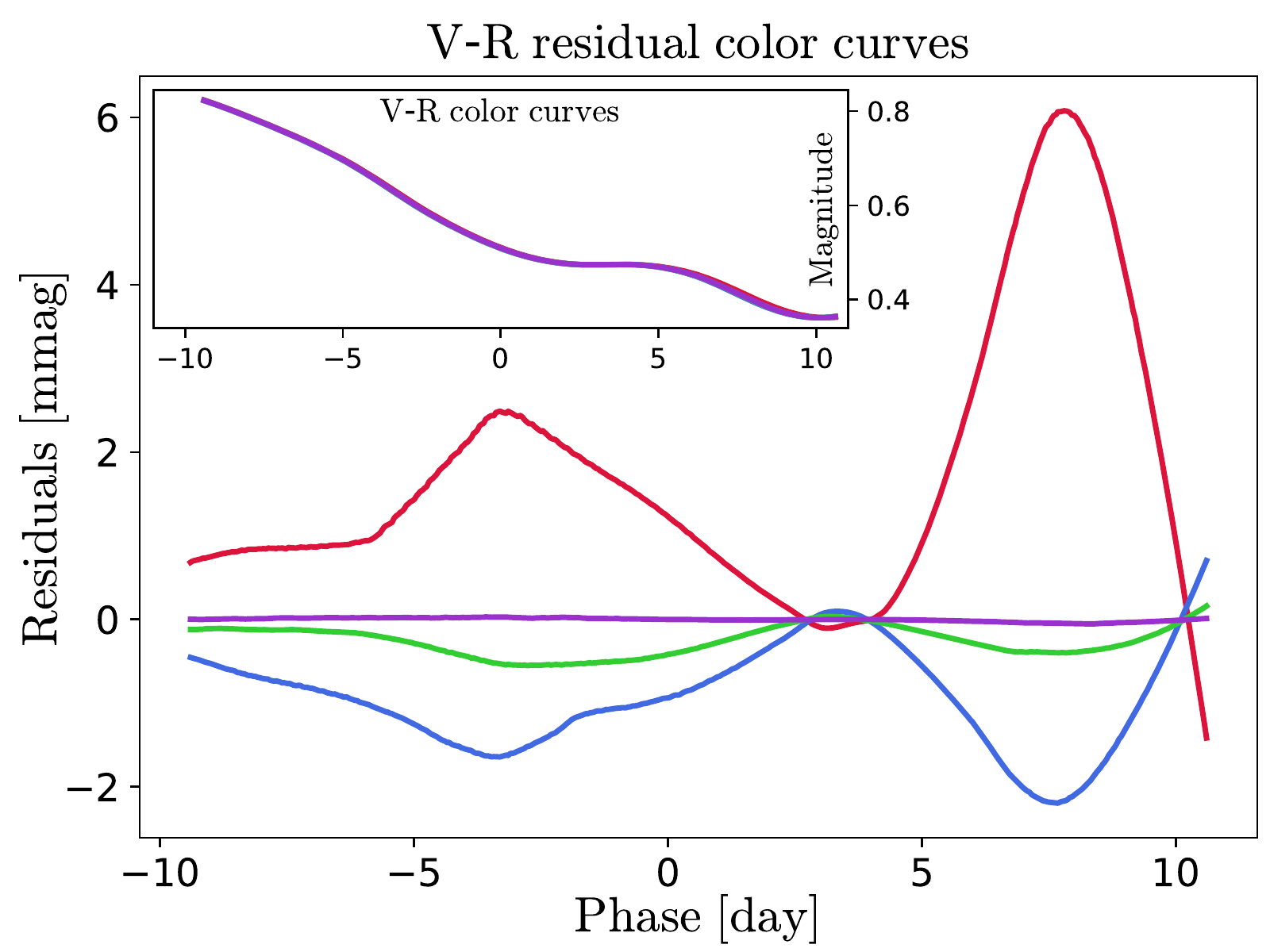}
    \end{minipage}    
        
    \caption{\textit{Left:} Toy model light curves for the single image D. The curves have been shifted in magnitude for visual purposes. The vertical ticks mark the observed peak luminosity. The black and purple curves overlap almost perfectly. \textit{Right:} ùResiduals of the microlensed V-R color curves with respect to the nonmicrolensed case. The corresponding position of the source in the microlensing map is indicated by colored circles in the left panel of Fig.~\ref{fig:mltd_evol}. The insert shows the V-R color curves prior to the subtraction of the non microlensed color curves. All the curves are displayed in the supernova reference frame, i.e., there is no time rescaling due to the redshift of the supernova.}
    \label{fig:lcs}
\end{figure*}

\subsection{Illustration with single band and color light curves}

To mitigate the impact of microlensing on time-delay measurements, G18 propose to use color curves. During the achromatic microlensing phase described in G18, microlensing magnification cancels out in color curves, easing the time-delay measurement with template fitting compared with single-band light curves. Marginalizing over various lens galaxy configurations and magnification patterns, G18 showed that one can recover the cosmological time delay up to an accuracy of $\sim 0.1$ days using color light curves. G18 simulations include by construction the geometric and microlensing time delay, although blended with other effects (their SNeIa model does not predict perfect achromaticity), which prevents a detailed analysis of the error budget of the time-delay measurement. Their precision is however consistent with our results presented in Table~\ref{tab:tab}, in which we formalise and isolate the impact of microlensing time delay alone, thus providing a lower limit in precision with which the cosmological time delays can be measured. The comparison between G18 accuracy and the present work is meaningful since our microlensing time-delay estimates are independent of the chosen filter, and are therefore also   valid for color curves, that is, the ratio of two single-band light curves. We assess this statement by creating V-R color curves in our toy model and computing the probability density of the mean excess of microlensing time-delay distributions. The change with respect to the values obtained with the R band light curves presented in Table~\ref{tab:tab} is again of the order of $10^{-6}$ days, that is, well below sub-percent. This shows that microlensing time delay is still present in color curves, although our toy model assumes a perfectly achromatic microlensing magnification.

To better illustrate how microlensing time delay impacts the light curves, as well as disentangle the contributions from microlensing magnification, geometrical time delay and microlensing time delay, we present in this section mock R-band light curves and R-V color light curves. We produce mock light curves for our toy model by computing the observed luminosity using Eq.~\ref{eq:flux}. The left panel of Fig. \ref{fig:lcs} presents R-band light curves for image D of iPTF16geu, in various configurations: i) the source is point-like and there is no microlensing (gray curve), ii) the source is an expanding shell and there is no microlensing (geometrical time delay only, black curve), and iii) the source is an expanding shell and there is microlensing (geometric and microlensing time delay as well as microlensing magnification, colored curves). The curves have been arbitrarily shifted in magnitude in order to match the same luminosity at phase $t=-10$ days. The difference between the maximum luminosity of the gray (point source) and black (no microlensing) curves illustrates the excess of geometrical time delay. We note that the excess of geometrical time delay is negative in this illustration (the maximum luminosity occurs earlier) due to our definition of the delay (see Eq.~\ref{eq:geometrical_delay}), in which the zero-delay point corresponds to the plane perpendicular to the line of sight and passing through the center of the supernova. The variation of the maximum luminosity in color curves with respect to the black curves is the combined effect of microlensing time delay and microlensing magnification, the latter being the dominant factor. 

The right panel of Fig.~\ref{fig:lcs} presents the residuals of the microlensed V-R color light curves after subtraction of the unmicrolensed color curve. The corresponding color curves are presented in the top-left insert. The plot shows that the residuals, although being very small (a few milli-magnitudes, only approximately $100$ times less than the residuals of single-band light curves), are nonzero. This indicates that the effect of microlensing does not completely cancel out in color curves. Microlensing time delay distorts in time the observed single band and color curves; the features of the color curves (maximas, minimas, etc.) appear as if shifted in time, thus no longer matching the corresponding features in the unmicrolensed color curve.

In order to assess the validity of the microlensing time delays computed through Eq.~\ref{eq:mldelay}, we compare the predictions of Table~\ref{tab:tab} with measurements performed on mock color light curves.The color curves are by construction free of microlensing magnification, and comparing the time shift of an event in the microlensed color curve with respect to its unmicrolensed equivalent allows for removal of the geometrical time-delay contribution. Doing so, we are left with microlensing time delay only. We generate 2000 sets of mock V-R light curves for image D, varying the source position in the microlensing maps for each set. The event chosen to estimate the microlensing time delay is the time coordinate of the minimum magnitude of the color light curves, occurring around $t\simeq10$ days in the rest frame (see insert of Fig.~\ref{fig:lcs}). The distribution of measured time shifts should follow the predicted microlensing time delay distribution, although a small deviation is expected; indeed, microlensing time delay not only produces a time shift of the characteristic features of the light curves, but also skews the overall shape of the light curves due to the reweighting of the source profile (see Eq.~\ref{eq:flux}), which affects the time coordinate of the minimum magnitude. Nevertheless, we find a measured shift of $-0.001^{+0.041}_{-0.059}$ days (50th, 16th and 84th percentiles of the distribution) in reasonably good agreement with the predicted microlensing time delay of image $D$ at phase $t=10$ days.

%===================
\section{Conclusions}
%===================

% final remarks, comparison with quasars
While microlensing time delay in lensed SNeIa is, on average, negligible with respect to the precision on time-delay measurements currently required for time-delay cosmography, it cannot always be ignored. For peculiar configurations, the bias can be of several tenths of a day, that is, enough to prevent sub-percent \hc\ determination from a small number of systems, especially if the cosmological delays are short. This is precisely the case of the only resolved lensed SNeIa discovered to date, iPTF16geu, for which the delays are predicted to be smaller than a day. The microlensing time delay computed in this work is unfortunately large enough to hamper the use of iPTF16geu for precision time-delay cosmography.

The careful and topical reader might wonder why the amplitude of microlensing time delay varies so much between lensed SNeIa and quasars \citep[up to several days; see TK18 and][]{Bonvin2018}. The answer resides in their spatial extent. Whereas SNeIa sizes in this work are of the order of light days, the thin-disk model for quasars extends to hundreds of light days. The bulk of the microlensing time-delay effect in quasars comes from the extended regions very far away from the center. However, TK18 do not consider any truncation radius in their model which, in real quasars, must occur somewhere. Depending on the value of this radius, the impact of microlensing time delay in lensed quasars may be smaller than claimed in TK18. For lensed supernovae, this truncation radius is naturally present, as the edge of the supernovae envelopes are sharp rather than exponentially decreasing. This is mostly why time delay microlensing seems, so far, less pronounced in supernovae than in quasars.

Our toy model includes simplifications but is sufficient for our purpose of illustrating the potential impact of geometrical microlensing time delay on lensed supernovae cosmography. Full 3D state-of-the-art radiative-transfer codes like \texttt{ARTIS} \citep{Kromer2009} or \texttt{SEDONA} \citep{Kasen2006} show a much more complex picture. In our toy model, we associate to each line-of-sight a single microlensing time delay, $\tau^{\mathrm{m}}$, whose value is computed at the intersect of the line-of-sight with the surface of the supernova. In reality, especially in the optically thin phase of the explosion, we should do the integration all along the supernova depth. As geometrical time delays associated with deeper layers of supernovae are also smaller, the net effect of our approximation is that we overestimate the amplitude of the microlensing time delay. This is ultimately good news for cosmological applications of lensed supernovae. Future work to better constrain microlensing time delay should focus on a more thorough modeling of the spatial emission of the source, especially during the optically thin phase.

%when the spatial extent of the emission region can reach hundreds of light days.

\begin{acknowledgements}
We thank the anonymous referee for his/her very useful comments, which greatly helped to improve the quality of the manuscript. This work is supported by the Swiss National Foundation. We thank Sherry Suyu, Markus Kromer, Ulrich N\"obauer and Cyril Georgy for useful comments and discussions. This research made use of Astropy, a community-developed core Python package for astronomy \citep{Astropy2013, Astropy2018} and the 2D graphics environment Matplotlib \citep{Hunter2007}.
\end{acknowledgements}

% \appendix
% \section{1st order correction in the case of relativistic expanding shell : }
% \warning{Martin : Some part of this should go in the Formalism section. But I put it here waiting for confirmation :}
% \com{Fred}{I would not put an appendix in such a short paper...}
% The relation between the 3D radius $\rho (t)$ and the radius observed in the plane of the sky $R_0(t)$ can be expressed to the first order as : 

% \begin{equation}
% R_0(t) = \rho(t) - \frac{d\rho(t)}{dt}\cdot \frac{\rho(t)}{c}
% \end{equation}
% We see immediatly that the two raidii are equal if the shell expands at a non-relativistic velocity : 
% \begin{equation}
% \frac{d\rho(t)}{dt} \ll c.
% \end{equation}
% The geometrical delay (Equations \ref{geo_delay}) can be re-expressed in the relativistic case keeping only the first order term : 
% \begin{equation}
%  \tau_{rel} (x,y,t) = \frac{\sqrt{r(x,y,t)^2-x^2-y^2}}{c}, 
% \end{equation}
% with : 
% \begin{equation}
% r(x,y,t) = \rho(t) - \frac{d\rho(t)}{dt}\left(\frac{\rho(t) - \sqrt{\rho(t)^2 - x^2 -y^2}}{c} \right)
% \end{equation}
% \warning{Actually, this is limited to first order, we should have $r(t)^2$ under the square root instead of $\rho(t)^2$ and this is recursive...}

\bibliographystyle{aa}
\bibliography{paper}

\end{document}